\documentclass[aps,twocolumn,superscriptaddress,preprintnumbers,amsmath,amssymb]{revtex4}
\usepackage{epsfig}
\usepackage{dcolumn}
\usepackage{bm}
\usepackage{amsmath}
\usepackage{color}
\usepackage[colorlinks,linkcolor=red,anchorcolor=green,citecolor=blue]{hyperref}
\usepackage{multirow}
\usepackage{relsize}

\newcommand{\BR}{{\cal B}}

\newcommand{\EE}{e^+e^-}

\newcommand{\gev}{\rm GeV}

\newcommand{\beq}{\begin{equation}}
\newcommand{\eeq}{\end{equation}}
\newcommand{\bitm}{\begin{itemize}}
\newcommand{\eitm}{\end{itemize}}

\newcommand{\add}[1]{{\color{blue}{#1}}}

\begin{document}

\title{\quad\\[0.2cm] Search for the $\eta_{c2}(1D)$ in $e^+e^-\to\gamma\eta_{c2}(1D)$ at $\sqrt{s}$ near 10.6 GeV at Belle}


\noaffiliation
\affiliation{Department of Physics, University of the Basque Country UPV/EHU, 48080 Bilbao}
\affiliation{University of Bonn, 53115 Bonn}
\affiliation{Brookhaven National Laboratory, Upton, New York 11973}
\affiliation{Budker Institute of Nuclear Physics SB RAS, Novosibirsk 630090}
\affiliation{Faculty of Mathematics and Physics, Charles University, 121 16 Prague}
\affiliation{Chonnam National University, Gwangju 61186}
\affiliation{University of Cincinnati, Cincinnati, Ohio 45221}
\affiliation{Deutsches Electronen-Synchrotron, 22607 Hamburg}
\affiliation{University of Florida, Gainesville, Florida 32611}
\affiliation{Department of Physics, Fu Jen Catholic University, Taipei 24205}
\affiliation{Key Laboratory of Nuclear Physics and Ion-beam Application (MOE) and Institute of Modern Physics, Fudan University, Shanghai 200443}
\affiliation{Justus-Liebig-Universit\"at Gie\ss{}en, 35392 Gie\ss{}en}
\affiliation{Gifu University, Gifu 501-1193}
\affiliation{SOKENDAI (The Graduate University for Advanced Studies), Hayama 240-0193}
\affiliation{Gyeongsang National University, Jinju 52828}
\affiliation{Department of Physics and Institute of Natural Sciences, Hanyang University, Seoul 04763}
\affiliation{University of Hawaii, Honolulu, Hawaii 96822}
\affiliation{High Energy Accelerator Research Organization (KEK), Tsukuba 305-0801}
\affiliation{J-PARC Branch, KEK Theory Center, High Energy Accelerator Research Organization (KEK), Tsukuba 305-0801}
\affiliation{National Research University Higher School of Economics, Moscow 101000}
\affiliation{Forschungszentrum J\"{u}lich, 52425 J\"{u}lich}
\affiliation{IKERBASQUE, Basque Foundation for Science, 48013 Bilbao}
\affiliation{Indian Institute of Science Education and Research Mohali, SAS Nagar, 140306}
\affiliation{Indian Institute of Technology Guwahati, Assam 781039}
\affiliation{Indian Institute of Technology Hyderabad, Telangana 502285}
\affiliation{Indian Institute of Technology Madras, Chennai 600036}
\affiliation{Indiana University, Bloomington, Indiana 47408}
\affiliation{Institute of High Energy Physics, Chinese Academy of Sciences, Beijing 100049}
\affiliation{Institute of High Energy Physics, Vienna 1050}
\affiliation{Institute for High Energy Physics, Protvino 142281}
\affiliation{INFN - Sezione di Napoli, I-80126 Napoli}
\affiliation{INFN - Sezione di Roma Tre, I-00146 Roma}
\affiliation{INFN - Sezione di Torino, I-10125 Torino}
\affiliation{Advanced Science Research Center, Japan Atomic Energy Agency, Naka 319-1195}
\affiliation{J. Stefan Institute, 1000 Ljubljana}
\affiliation{Institut f\"ur Experimentelle Teilchenphysik, Karlsruher Institut f\"ur Technologie, 76131 Karlsruhe}
\affiliation{Department of Physics, Faculty of Science, King Abdulaziz University, Jeddah 21589}
\affiliation{Kitasato University, Sagamihara 252-0373}
\affiliation{Korea Institute of Science and Technology Information, Daejeon 34141}
\affiliation{Korea University, Seoul 02841}
\affiliation{Kyoto Sangyo University, Kyoto 603-8555}
\affiliation{Kyungpook National University, Daegu 41566}
\affiliation{Universit\'{e} Paris-Saclay, CNRS/IN2P3, IJCLab, 91405 Orsay}
\affiliation{P.N. Lebedev Physical Institute of the Russian Academy of Sciences, Moscow 119991}
\affiliation{Liaoning Normal University, Dalian 116029}
\affiliation{Faculty of Mathematics and Physics, University of Ljubljana, 1000 Ljubljana}
\affiliation{Ludwig Maximilians University, 80539 Munich}
\affiliation{Luther College, Decorah, Iowa 52101}
\affiliation{Malaviya National Institute of Technology Jaipur, Jaipur 302017}
\affiliation{Faculty of Chemistry and Chemical Engineering, University of Maribor, 2000 Maribor}
\affiliation{Max-Planck-Institut f\"ur Physik, 80805 M\"unchen}
\affiliation{School of Physics, University of Melbourne, Victoria 3010}
\affiliation{University of Mississippi, University, Mississippi 38677}
\affiliation{University of Miyazaki, Miyazaki 889-2192}
\affiliation{Moscow Physical Engineering Institute, Moscow 115409}
\affiliation{Graduate School of Science, Nagoya University, Nagoya 464-8602}
\affiliation{Universit\`{a} di Napoli Federico II, I-80126 Napoli}
\affiliation{Nara Women's University, Nara 630-8506}
\affiliation{National Central University, Chung-li 32054}
\affiliation{National United University, Miao Li 36003}
\affiliation{Department of Physics, National Taiwan University, Taipei 10617}
\affiliation{H. Niewodniczanski Institute of Nuclear Physics, Krakow 31-342}
\affiliation{Nippon Dental University, Niigata 951-8580}
\affiliation{Niigata University, Niigata 950-2181}
\affiliation{Novosibirsk State University, Novosibirsk 630090}
\affiliation{Osaka City University, Osaka 558-8585}
\affiliation{Pacific Northwest National Laboratory, Richland, Washington 99352}
\affiliation{Panjab University, Chandigarh 160014}
\affiliation{Peking University, Beijing 100871}
\affiliation{University of Pittsburgh, Pittsburgh, Pennsylvania 15260}
\affiliation{Research Center for Nuclear Physics, Osaka University, Osaka 567-0047}
\affiliation{Meson Science Laboratory, Cluster for Pioneering Research, RIKEN, Saitama 351-0198}
\affiliation{Department of Modern Physics and State Key Laboratory of Particle Detection and Electronics, University of Science and Technology of China, Hefei 230026}
\affiliation{Seoul National University, Seoul 08826}
\affiliation{Showa Pharmaceutical University, Tokyo 194-8543}
\affiliation{Soochow University, Suzhou 215006}
\affiliation{Soongsil University, Seoul 06978}
\affiliation{Sungkyunkwan University, Suwon 16419}
\affiliation{School of Physics, University of Sydney, New South Wales 2006}
\affiliation{Department of Physics, Faculty of Science, University of Tabuk, Tabuk 71451}
\affiliation{Tata Institute of Fundamental Research, Mumbai 400005}
\affiliation{Department of Physics, Technische Universit\"at M\"unchen, 85748 Garching}
\affiliation{Toho University, Funabashi 274-8510}
\affiliation{Department of Physics, Tohoku University, Sendai 980-8578}
\affiliation{Earthquake Research Institute, University of Tokyo, Tokyo 113-0032}
\affiliation{Department of Physics, University of Tokyo, Tokyo 113-0033}
\affiliation{Virginia Polytechnic Institute and State University, Blacksburg, Virginia 24061}
\affiliation{Wayne State University, Detroit, Michigan 48202}
\affiliation{Yamagata University, Yamagata 990-8560}
\affiliation{Yonsei University, Seoul 03722}
  \author{S.~Jia}\affiliation{Key Laboratory of Nuclear Physics and Ion-beam Application (MOE) and Institute of Modern Physics, Fudan University, Shanghai 200443} 
  \author{C.~P.~Shen}\affiliation{Key Laboratory of Nuclear Physics and Ion-beam Application (MOE) and Institute of Modern Physics, Fudan University, Shanghai 200443} 
  \author{I.~Adachi}\affiliation{High Energy Accelerator Research Organization (KEK), Tsukuba 305-0801}\affiliation{SOKENDAI (The Graduate University for Advanced Studies), Hayama 240-0193} 
  \author{H.~Aihara}\affiliation{Department of Physics, University of Tokyo, Tokyo 113-0033} 
  \author{S.~Al~Said}\affiliation{Department of Physics, Faculty of Science, University of Tabuk, Tabuk 71451}\affiliation{Department of Physics, Faculty of Science, King Abdulaziz University, Jeddah 21589} 
  \author{D.~M.~Asner}\affiliation{Brookhaven National Laboratory, Upton, New York 11973} 
  \author{T.~Aushev}\affiliation{National Research University Higher School of Economics, Moscow 101000} 
  \author{R.~Ayad}\affiliation{Department of Physics, Faculty of Science, University of Tabuk, Tabuk 71451} 
  \author{V.~Babu}\affiliation{Deutsches Electronen-Synchrotron, 22607 Hamburg} 
  \author{P.~Behera}\affiliation{Indian Institute of Technology Madras, Chennai 600036} 
% ...
\collaboration{The Belle Collaboration}

\begin{abstract}

For the first time we search for the $\eta_{c2}(1D)$ in $e^+e^-\to\gamma\eta_{c2}(1D)$ at $\sqrt{s}$ = 10.52, 10.58, and 10.867 GeV with data
samples of 89.5 fb$^{-1}$, 711 fb$^{-1}$, and 121.4 fb$^{-1}$, respectively, accumulated with the Belle detector
at the KEKB asymmetric energy electron-positron collider. No significant $\eta_{c2}(1D)$ signal is observed in the mass range between 3.8 and 3.88 GeV/$c^2$. The upper limit at 90\% confidence level on the product of the Born cross section for $e^+e^- \to \gamma\eta_{c2}(1D)$ and branching fraction for $\eta_{c2}(1D)\to \gamma h_c(1P)$ is determined to be $\sigma(e^+e^- \to \gamma\eta_{c2}(1D))\BR(\eta_{c2}(1D)\to \gamma h_c(1P))$ $<$ 4.9 fb at $\sqrt{s}$ near 10.6 GeV.

\end{abstract}

\maketitle

The contemporary interest into heavy quarkonia largely stems from the fact that many experimental results are poorly understood on the theoretical side.
Discoveries of new states or new processes, and improved precisions for known states in experiments help to verify various QCD models. Among the charmonium states, a $D$-wave charmonium state $\psi_2(1D)$ decaying into $\gamma\chi_{c1}$ was observed in $B$ decays and $e^+e^-$ annihilations by Belle~\cite{032001} and BESIII~\cite{011803}, respectively. The mass of $\psi_2(1D)$ is $(3822.2\pm1.2)$ MeV/$c^2$~\cite{032001,011803}, and the upper limit at 90\% confidence level (C.L.) on the width is 16 MeV~\cite{011803}. Using proton-proton collision data, LHCb reported the observation of $X(3842)$ in $D^0\bar D^0$ and $D^+D^-$ mass spectra~\cite{035}. The observed mass and width are $(3842.71\pm0.16\pm0.12)$ MeV/$c^2$ and $(2.79\pm0.51\pm0.35)$ MeV, which suggest the interpretation of the new state as the spin-3 charmonium $\psi_3(1D)$ state~\cite{035}.~Only for the spin-singlet low-lying $D$-wave state $\eta_{c2}(1D)$ there has been no experimental clue yet.

The mass of $\eta_{c2}(1D)$ is predicted to be in the range of 3.80 to 3.88 GeV/$c^2$ by various potential models~\cite{189,2079,5229,014027,094019,054026,094004}, which lies between the $D\bar D$ and $D^*\bar D$ thresholds. Lattice calculations~\cite{034501,089} find that the $\eta_{c2}(1D)$ and $\psi_2(1D)$ masses are close to each other. Considering the hyperfine splitting of the $1D$ charmonium states is expected to be small, we can also deduce the mass of $\eta_{c2}(1D)$ using the known masses of the $1^3D_J$ states: $m_{\eta_{c2}(1D)}$ $\approx$ $(3m_{\psi(3770)}+5m_{\psi_2(1D)}+7m_{X(3842)})/15$ $\approx$ 3822 MeV/$c^2$~\cite{034}. Quite different from $\psi(3770)$, the decay of $\eta_{c2}(1D)$ into $D\bar D$ is forbidden due to the conservation of parity. Thus $\eta_{c2}(1D)$ is a narrow resonance, and its main decay modes are considered to be hadronic decays and electromagnetic E1 transitions. The branching fraction of $\eta_{c2}(1D)\to \gamma h_c(1P)$ is  larger than 50\% over a large number of different predictions~\cite{014027,054026,162002,014001}.

The first search for the $\eta_{c2}(1D)$ was carried out in $B$ decays by Belle based on the 711 fb$^{-1}$ data sample collected on the $\Upsilon(4S)$ resonance~\cite{034}. The decays of $B^+\to \eta_{c2}(1D)K^+$, $B^0\to \eta_{c2}(1D)K^0_S$, $B^0\to \eta_{c2}(1D)\pi^-K^+$, and $B^+\to \eta_{c2}(1D)\pi^+K^0_S$ with $\eta_{c2}(1D)\to \gamma h_c(1P)$ were extensively investigated. No significant $\eta_{c2}(1D)$ signals were observed. The upper limits at 90\% C.L. on the product of the branching fractions ($\BR(B\to \eta_{c2}(1D) + h)\BR(\eta_{c2}(1D)\to \gamma h_c(1P)),~h=K~{\rm or}~K\pi$) are at the level of $10^{-5}$ to $10^{-4}$~\cite{034}. Searches for conventional charmonia in the reactions $e^+e^- \to \gamma \chi_{cJ}$ ($J$ = 0, 1, 2) and $\gamma\eta_c(1S)$ have been performed by Belle at center-of-mass energies (C.M.) 10.52, 10.58, and 10.867 GeV, respectively~\cite{092015}. A significant $\gamma\chi_{c1}$ signal was observed for the first time at $\sqrt{s}$ = 10.58 GeV with a significance of 5.1$\sigma$ including systematic uncertainties. The reported Born cross section is relatively large; $\sigma(e^+e^-\to \gamma \chi_{c1})$ = $17.3^{+4.2}_{-3.9}({\rm stat.})\pm1.7({\rm syst.})$ fb~\cite{092015}. This may indicate a large production rate for other conventional charmonia, i.e. $\eta_{c2}(1D)$, in $e^+e^- \to \gamma\eta_{c2}(1D)$. From theory, the cross section of $e^+e^-\to\gamma\eta_{c2}(1D)$ at $\sqrt{s}$ = 10.58 GeV is calculated to be 1.5 fb within the framework of non-relativistic QCD factorization formalism~\cite{14364}.

In this paper, we search for the $\eta_{c2}(1D)$ in $e^+e^- \to \gamma\eta_{c2}(1D)$ at $\sqrt{s}$ $\sim$ 10.6 GeV with a data sample of 921.9 fb$^{-1}$ at Belle. We search for $\eta_{c2}(1D)$ decaying into $\gamma h_c(1P)$, and $h_c(1P)$ decaying into $\gamma \eta_c(1S)$. The $\eta_c(1S)$ candidates are reconstructed via five hadronic decays of $K^0_S(\to\pi^+\pi^-)K^+\pi^-$, $\pi^+\pi^-K^+K^-$, $2(\pi^+\pi^-)$, $2(K^+K^-)$, and $3(\pi^+\pi^-)$. For $\eta_c(1S)\to K^0_S K^+\pi^-$, the charge-conjugated mode is implied.

The data used in this analysis correspond to 89.5 fb$^{-1}$ of integrated luminosity at 10.52 GeV, 711 fb$^{-1}$ at 10.58 GeV, and 121.4 fb$^{-1}$ at 10.867 GeV, respectively, which were recorded by the Belle detector~\cite{Belle1,Belle2} at the KEKB asymmetric-energy $e^+e^-$ collider~\cite{KEKB1,KEKB2}.
The Belle detector was a large solid-angle magnetic spectrometer that consists of a silicon vertex detector, a 50-layer central drift chamber (CDC), an array of aerogel threshold Cherenkov counters (ACC), a barrel-like arrangement of time-of-flight scintillation counters (TOF), and an electromagnetic calorimeter comprised of CsI(TI) crystals (ECL) located inside a superconducting solenoid coil that provides a 1.5T magnetic field.
The ECL is divided into three regions spanning $\theta$, the angle of inclination in the laboratory frame with respect to the direction opposite the $e^+$ beam. The ECL backward end cap, barrel, and forward end cap cover ranges of $-0.91 < \cos\theta < -0.65$, $-0.63 < \cos\theta < 0.85$, and $0.85 < \cos\theta < 0.98$, respectively. An iron flux-return yoke instrumented with resistive plate chambers located outside the coil was used to detect $K^0_L$ mesons and identify muons.

Monte Carlo (MC) samples are generated with {\sc evtgen}~\cite{EVTGEN} to determine detection efficiency and optimize event selection criteria. The simulated events are processed with a detector simulation based on {\sc geant3}~\cite{geant3}. No definite model exists for the distribution of polar angle $\theta_\gamma$ in the $e^+e^-$ C.M. system for $e^+e^- \to \gamma \eta_{c2}(1D)$ because the combination of tensor-meson production and $\gamma$ emission is theoretically complicated and requires experimental input. So we model the production as evenly distributed in phase space and account for differences from $(1\pm\cos^2\theta_\gamma)$ distributions as systematic uncertainties. Correction of initial state radiation (ISR) is taken into account in the studied mode, where we assume the Born cross section $\sigma(e^+e^- \to \gamma \eta_{c2}(1D)) \sim 1/s^n$ ($n$ = 2) in the calculation of the radiative correction factor, where $s$ is the $e^+e^-$ C.M. energy squared. Changing the $s$ dependence of the cross section from $n$ = 2 to $n$ = 1 or $n$ = 4, the maximum difference of the radiative correction factor is considered as the systematic uncertainty. Generic MC samples, i.e.~$B$ = $B^+$, $B^0$, or $B^{(*)}_s$ decays and $e^+e^-\to q\bar q$ ($q$ = $u$, $d$, $s$, $c$) at $\sqrt{s}$ = 10.52, 10.58, and 10.867 GeV, normalized to the same integrated luminosity as real data, are used to check for possible peaking backgrounds~\cite{107540}.

Except for the charged tracks from the relatively long-lived $K^0_S$ decaying into $\pi^+\pi^-$, impact parameters with respect to the interaction point are required to be less than 0.5 cm perpendicular to and 2 cm along the beam axis, respectively.~For $e^+e^- \to \gamma \eta_{c2}(1D)$, we require the numbers of charged tracks, $N_{\rm trk}$, to be exactly 6 for $\eta_c(1S)\to 3(\pi^+\pi^-)$ and 4 for $\eta_c(1S)\to K^0_SK^+\pi^-/\pi^+\pi^-K^+K^-/2(\pi^+\pi^-)/2(K^+K^-)$, also with a zero net charge.
For the particle identification (PID) of a well-reconstructed charged track, information from different detector subsystems, including specific ionization in the CDC, time measurement in the TOF, and the response of the ACC, is combined to form a likelihood ${\cal L}_i$~\cite{PID} for particle species $i$. Tracks with $R_K={\cal L}_K/({\cal L}_K+{\cal L}_\pi)$ $<$ 0.4 are identified as pions with an efficiency of 93\%, while 6\% of kaons are misidentified as pions; tracks with $R_K$ $>$ 0.6 are identified as kaons with an efficiency of 94\%, while 7\% of pions are misidentified as kaons.

Using a multivariate analysis with a neural network~\cite{190} based on two sets of input variables~\cite{2014}, a $K^0_S$ candidate is reconstructed from a pair of oppositely charged tracks that are treated as pions.~The invariant mass of the $K^0_S$ candidates is required to be within 10 MeV/$c^2$ ($\sim2.5\sigma$) of the nominal $K^0_S$ mass~\cite{PDG}. The $\eta_c(1S)$ candidates are required to satisfy $|M(\eta_c(1S)) - m_{\eta_c(1S)}|$ $<$ 60 MeV/$c^2$ (see below), where $m_{\eta_c(1S)}$ is the nominal mass of $\eta_c(1S)$~\cite{PDG}.

An ECL cluster is treated as a photon candidate if it is isolated from the projected path of charged tracks in the CDC. The energy of photons is required to be larger than 50 MeV in the barrel and larger than 100 MeV in the endcaps. The ratio of energy deposited in the 3 × 3 array of crystals, with the center being the one with the largest energy deposit, to that in the 5 × 5 array sharing the same center is required to be greater than 0.8. The number of photons \add{in an event} is required to be at least 3. The most energetic photon in the C.M. frame ($\gamma_{\rm max}$) is regarded as the primary photon in the process $e^+ e^- \to \gamma \eta_{c2}(1D)$. Among the remaining photons, all combinations of two photons are considered for the decay chain of $\eta_{c2}(1D)$. We call the photon that appears directly from $\eta_{c2}(1D)$ decay as $\gamma_1$, and the one from $h_c(1P)$ as $\gamma_2$. We require $|M(\eta_c(1S)\gamma_2) - m_{h_c(1P)}|$ $<$ 30 MeV/$c^2$ ($\sim2.5\sigma$) and 3.7 $<$ $M(\eta_c(1S)\gamma_1\gamma_2)$ $<$ 4.0 GeV/$c^2$, where $m_{h_c(1P)}$ is the nominal mass of $h_c(1P)$~\cite{PDG}. Note that we do not distinguish $\gamma_1$ and $\gamma_2$, i.e. rank them by their energies, to increase efficiency and avoid possible bias. In this case, the incorrect attribution rate of $\gamma_1 \eta_c(1S)/\gamma_2 \eta_c(1S)$ for $h_c(1P)$ signal is only 0.8\% according to signal MC simulations after applying a five-constraint (5C) kinematic fit (see below) for final states from $\eta_c(1S)$, $\gamma_{\rm max}$, $\gamma_1$, and $\gamma_2$. In both signal MC and the data, no events have multiple entries. Therefore, the entire decay chain can be written as $e^+e^- \to \gamma_{\rm max}\eta_{c2}(1D)$, $\eta_{c2}(1D)\to \gamma_1 h_c(1P)$, $h_c(1P)\to\gamma_2\eta_c(1S)$, $\eta_c(1S)\to{\rm hadrons}$.

A 5C kinematic fit constraining the four-momenta of the final-state particles to the initial $e^+e^-$ collision system and the invariant mass of $\gamma_2\eta_c(1S)$ to the $h_c(1P)$ nominal mass~\cite{PDG} is performed; the fit has five degrees of freedom. The fit is required to satisfy $\chi^2_{\rm 5C}$ $<$ 20 to improve the resolutions of the momenta of charged tracks and the energies of photons, and to suppress backgrounds with more than three photons, such as ISR processes. The requirement of $\chi^2_{\rm 5C}$ is optimized by maximizing the Punzi parameter $S/(3/2+\sqrt{B})$~\cite{punzi}, where $S$ is the number of signal events in signal MC samples assuming $\sigma(e^+e^- \to \gamma\eta_{c2}(1D))=1.5$ fb~\cite{14364} and $\BR(\eta_{c2}(1D)\to \gamma h_c(1P))$ = 0.5~\cite{014027,054026,162002,014001}, and $B$ is the number of background events from $h_c(1P)$ mass sidebands in data (see below).

We veto $\gamma$ from $\pi^0$ and $\eta$ having $|M(\gamma_i\gamma_j)-m_{\pi^0}|$ $<$ 15 MeV/$c^2$ and $|M(\gamma_i\gamma_j)-m_{\eta}|$ $<$ 30 MeV/$c^2$, where $\gamma_i$ is either $\gamma_1$ or $\gamma_2$ and $\gamma_j$ is any photon in the event other than $\gamma_i$.
These veto ranges correspond to about 2.5$\sigma$ in resolution. The signal efficiencies are decreased by 1.5\% and 10.6\% after vetoing $\pi^0$ and $\eta$ backgrounds.

\begin{figure*}[htbp]
\centering
\includegraphics[width=3.7cm,angle=-90]{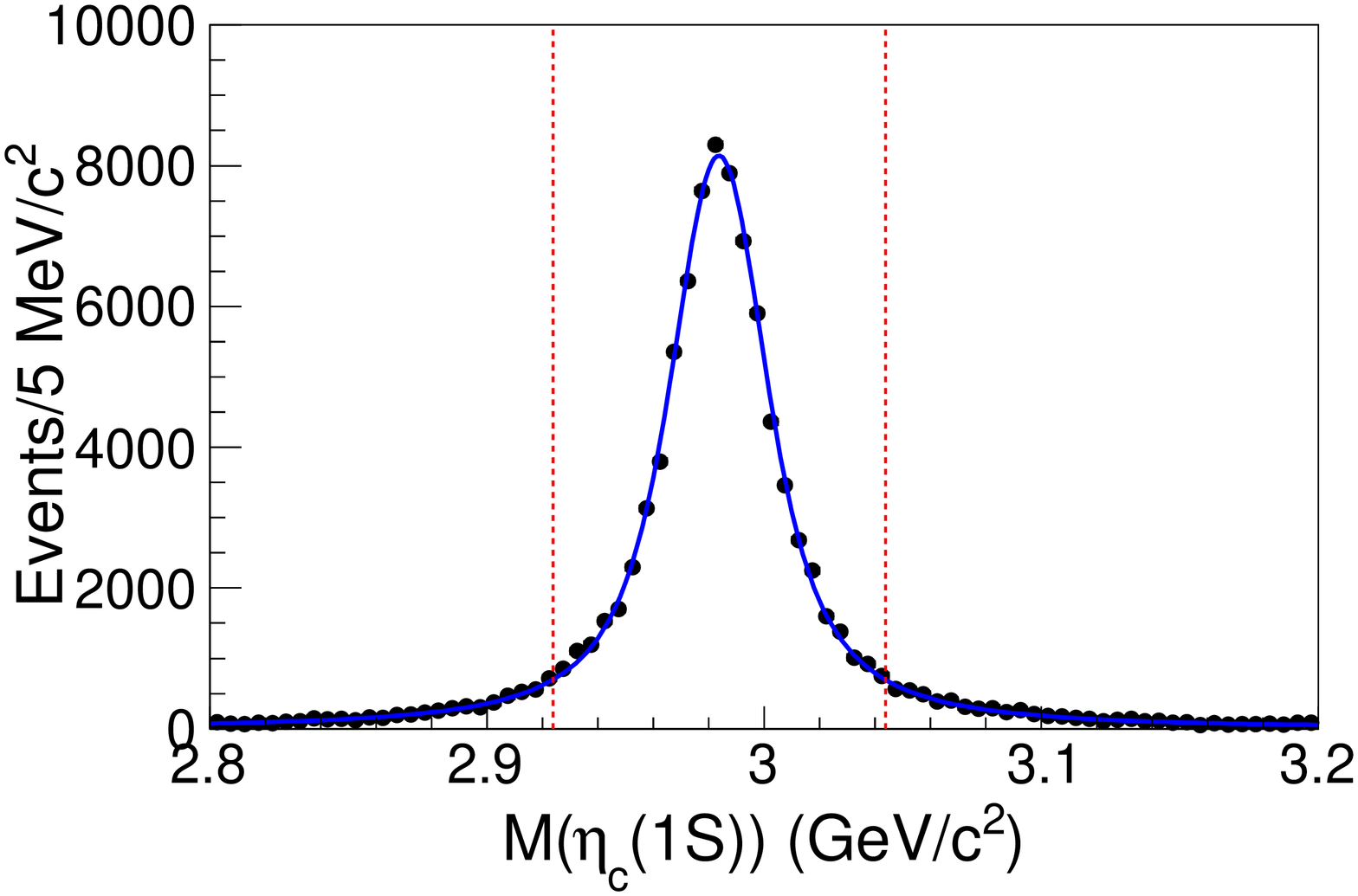}
\put(-30, -20){\large \bf (a)}
\hspace{0.10cm}
\includegraphics[width=3.7cm,angle=-90]{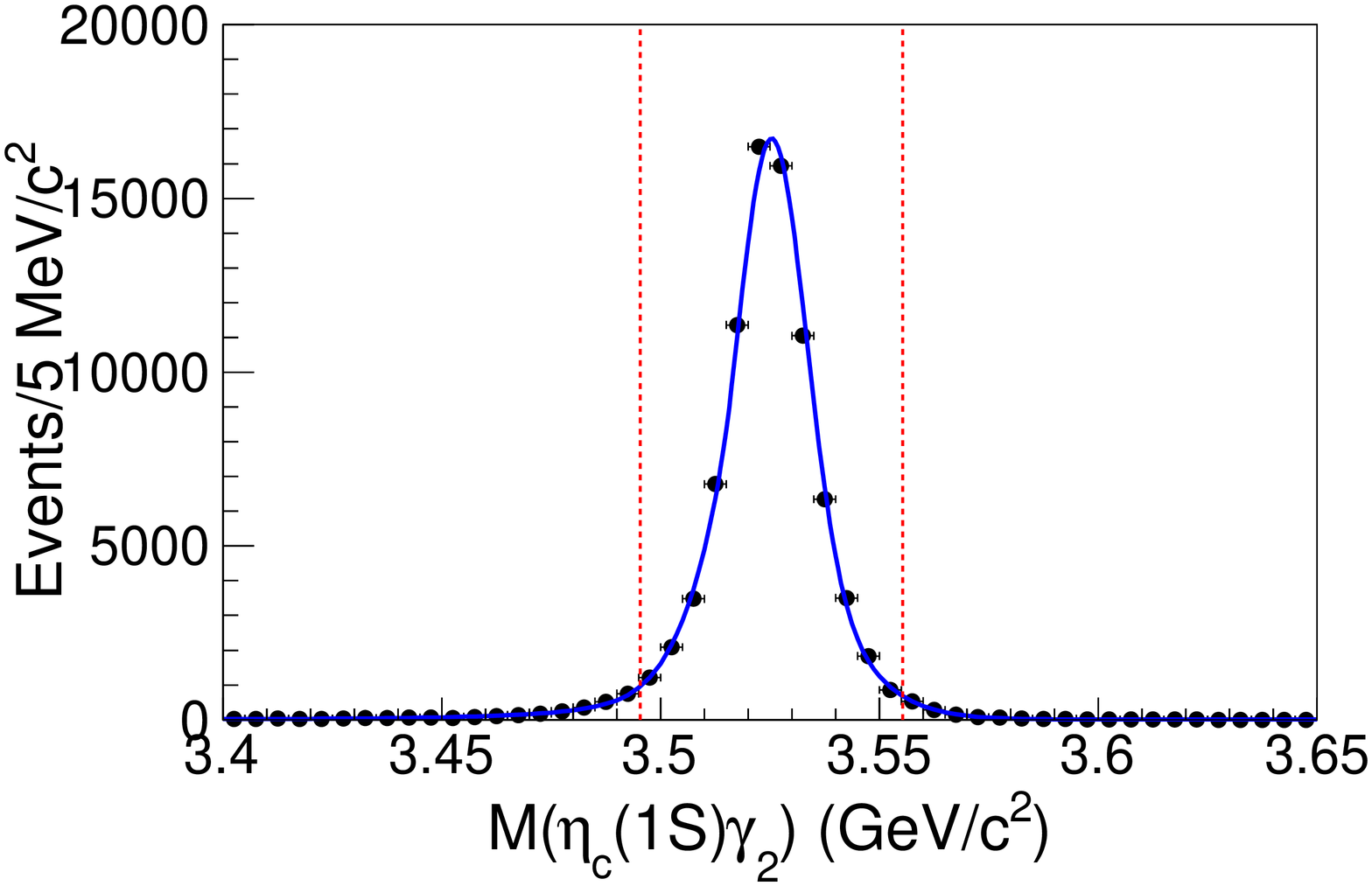}
\put(-30, -20){\large \bf (b)}
\hspace{0.10cm}
\includegraphics[width=3.7cm,angle=-90]{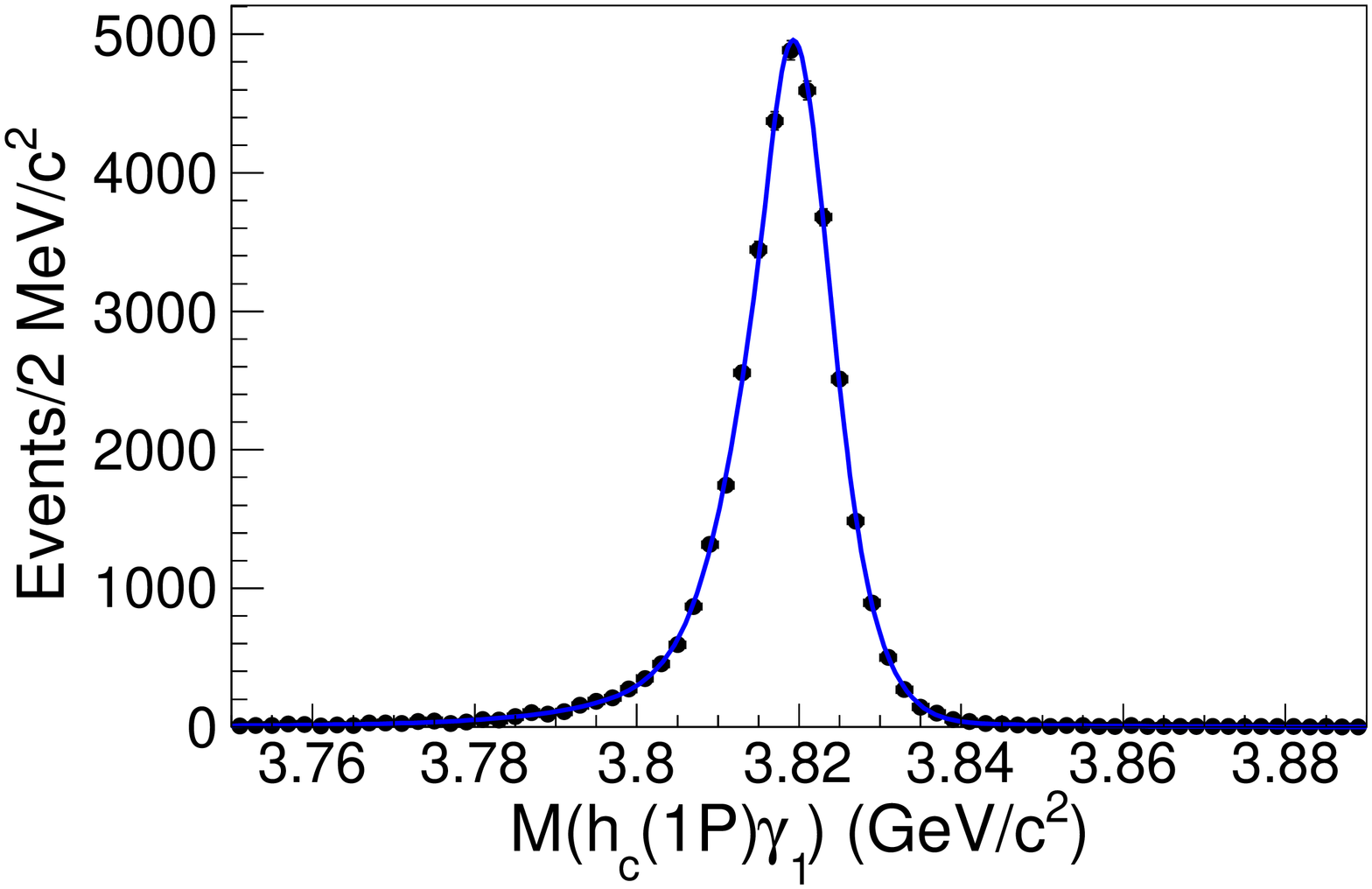}
\put(-28, -20){\large \bf (c)}
\caption{The invariant mass distributions for the selected (a) $\eta_c(1S)$, (b) $h_c(1P)$, and (c) $\eta_{c2}(1D)$ candidates from signal MC samples. The blue solid curves show the fitted results. The red dashed curves show the required signal regions.}\label{mc}
\end{figure*}

After applying the above requirements, the invariant mass distribution for the selected $\eta_c(1S)$ candidates from signal MC samples is shown in Fig.~\ref{mc}(a). The $\eta_c(1S)$ signal shape is described by a Breit-Wigner (BW) convolved with a Gaussian function. The fitted result is shown by the solid curve in Fig.~\ref{mc}(a). Here, we require $\eta_c(1S)$ candidates to satisfy $|M(\eta_c(1S)) - m_{\eta_c(1S)}|$ $<$ 60 MeV/$c^2$, where $m_{\eta_c(1S)}$ is the nominal mass of $\eta_c(1S)$~\cite{PDG}. In this region, more than 90\% signal events are retained. This relatively tight mass requirement is applied to avoid possible ISR backgrounds with $J/\psi$ candidates.

The invariant mass spectrum, before	the application of the 5C kinematic fit, for $h_c(1P)$ candidates from signal MC samples is shown in Fig.~\ref{mc}(b).
We fit the $M(\eta_c(1S)\gamma_2)$ distribution by modelling the $h_c(1P)$ signal with the convolution of a Crystal Ball (CB)~\cite{CB} and Gaussian function due to the asymmetric energy resolution of $\gamma_2$. The $h_c(1P)$ signal is required to satisfy $|M(\eta_c(1S)\gamma_2) - m_{h_c(1P)}|$ $<$ 30 MeV/$c^2$ ($\sim2.5\sigma$), where $m_{h_c(1P)}$ is the nominal mass of $h_c(1P)$~\cite{PDG}.

Figure~\ref{mc}(c) shows the invariant mass spectrum for $\eta_{c2}(1D)$ candidates in MC simulated $e^+e^-\to \gamma\eta_{c2}(1D)$ samples at $\sqrt{s}$ = 10.58 GeV with the $\eta_{c2}(1D)$ mass fixed at 3.82 GeV/$c^2$ and the width fixed at 0 MeV/$c^2$. The $\eta_{c2}(1D)$ signal shape is described by the convolution of a CB and Gaussian function. Based on the fitted results, we obtain the efficiencies for each $\eta_c(1S)$ decay mode ($\varepsilon_i$). Further, one can obtain the value of $\varepsilon_i\times\BR_i$ for each mode, where $\BR_i$ is the product of all secondary branching fractions~\cite{PDG}. The values of $\Sigma_i\varepsilon_i\BR_i$ at $\sqrt{s}$ = 10.52, 10.58, and 10.867 GeV are obtained to be $(20.4\pm2.0)\times10^{-4}$, $(20.3\pm2.0)\times10^{-4}$, and $(20.2\pm2.0)\times10^{-4}$, respectively, which are used to calculate the final Born cross section for $e^+e^-\to \gamma\eta_{c2}(1D)$. The value of $\Sigma_i\varepsilon_i\BR_i$ is independent of the mass of $\eta_{c2}(1D)$ in our studied region of [3.80, 3.88] GeV/$c^2$.

\begin{figure*}[htbp]
\centering
\includegraphics[width=4cm,angle=-90]{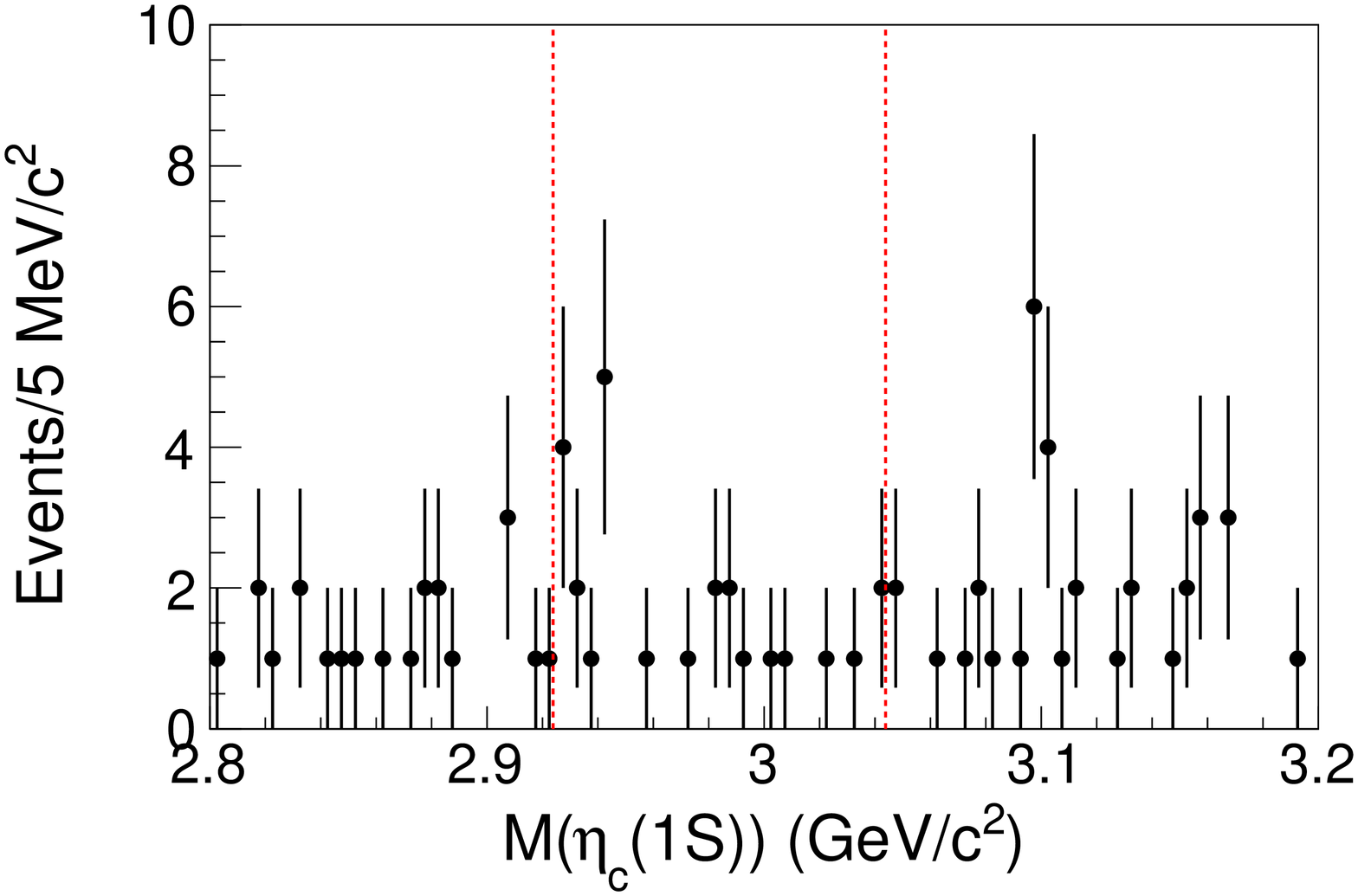}
\put(-30, -20){\large \bf (a)}
\hspace{0.4cm}
\includegraphics[width=4.cm,angle=-90]{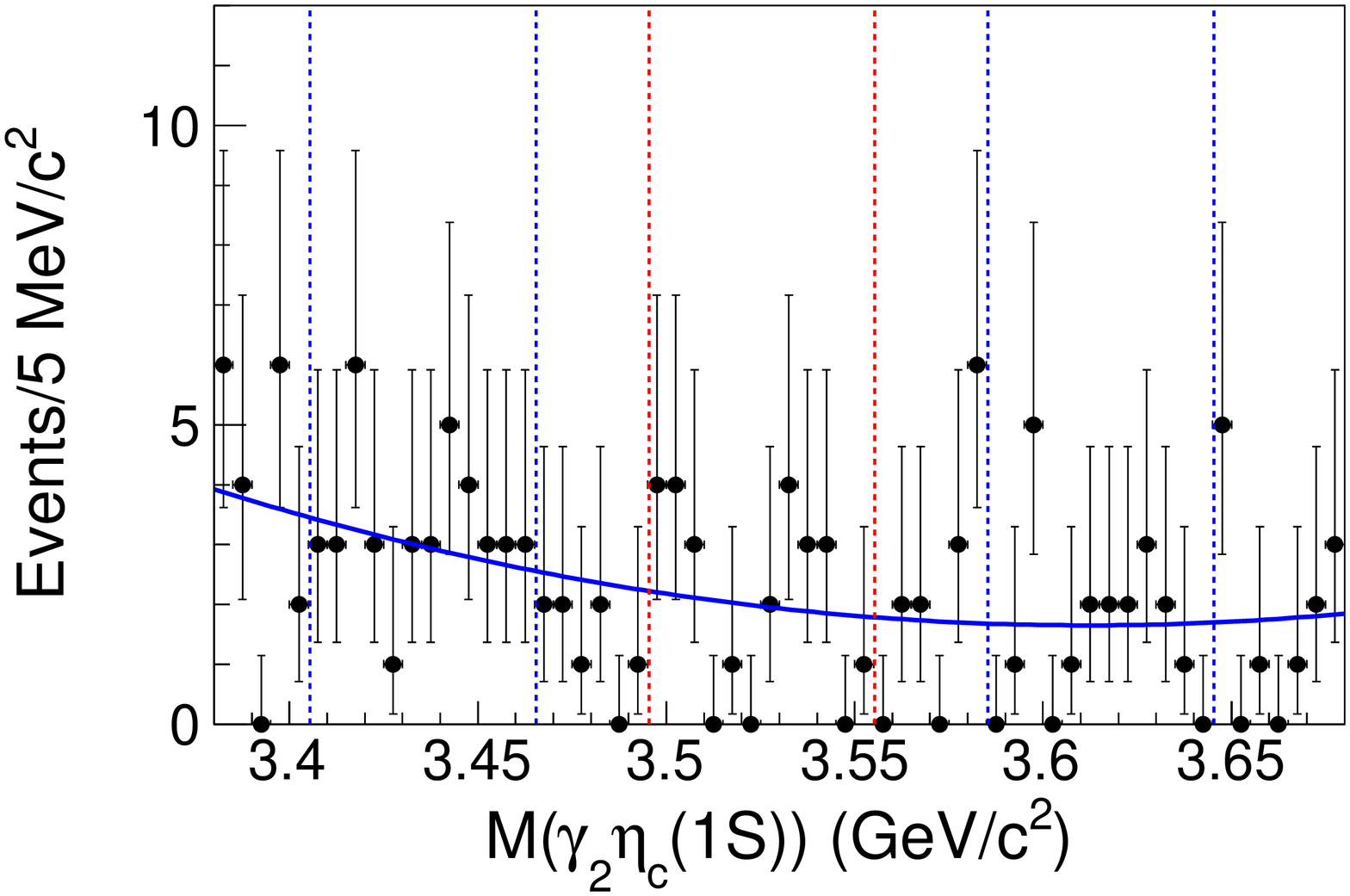}
\put(-35, -20){\large \bf (b)}
\caption{The invariant mass distributions for (a) $\eta_c(1S)$ and (b) $h_c(1P)$ candidates from a combined $\sqrt{s}$ = 10.52, 10.58, and 10.867 GeV data sample. The red dashed curves show the required signal regions. The blue solid curve in (b) shows the fitted result. The blue dashed curves show the defined $h_c(1P)$ mass sidebands.}\label{data}
\end{figure*}

After applying all the above requirements, the invariant mass distributions for $\eta_c(1S)$ and $h_c(1P)$ candidates from a combined $\sqrt{s}$ = 10.52, 10.58, and 10.867 GeV data sample are shown in Fig.~\ref{data}. No clear $\eta_c(1S)$ and $h_c(1P)$ signals can be seen. A few events are around the $J/\psi$ mass point~\cite{PDG}, where three candidates are expected from ISR production of $\psi(2S)$, with $\psi(2S) \to \gamma \chi_{c1} (\to \gamma J/\psi)$. The red dashed curves show the required signal regions as above. The blue dashed curves in Fig.~\ref{data}(b) \add{indicate} the $h_c(1P)$ mass sidebands on each side of the signal region. Each sideband is separated from the signal region with a 30 MeV/$c^2$ gap and has the same width as the signal region. To determine the normalization factor of $h_c(1P)$ mass sidebands, a second-order polynomial function is used to describe the $\gamma_2\eta_c(1S)$ invariant mass spectrum, as shown by the blue solid curve in Fig.~\ref{data}(b).

\begin{figure*}[htbp]
\centering
\includegraphics[width=3.7cm,angle=-90]{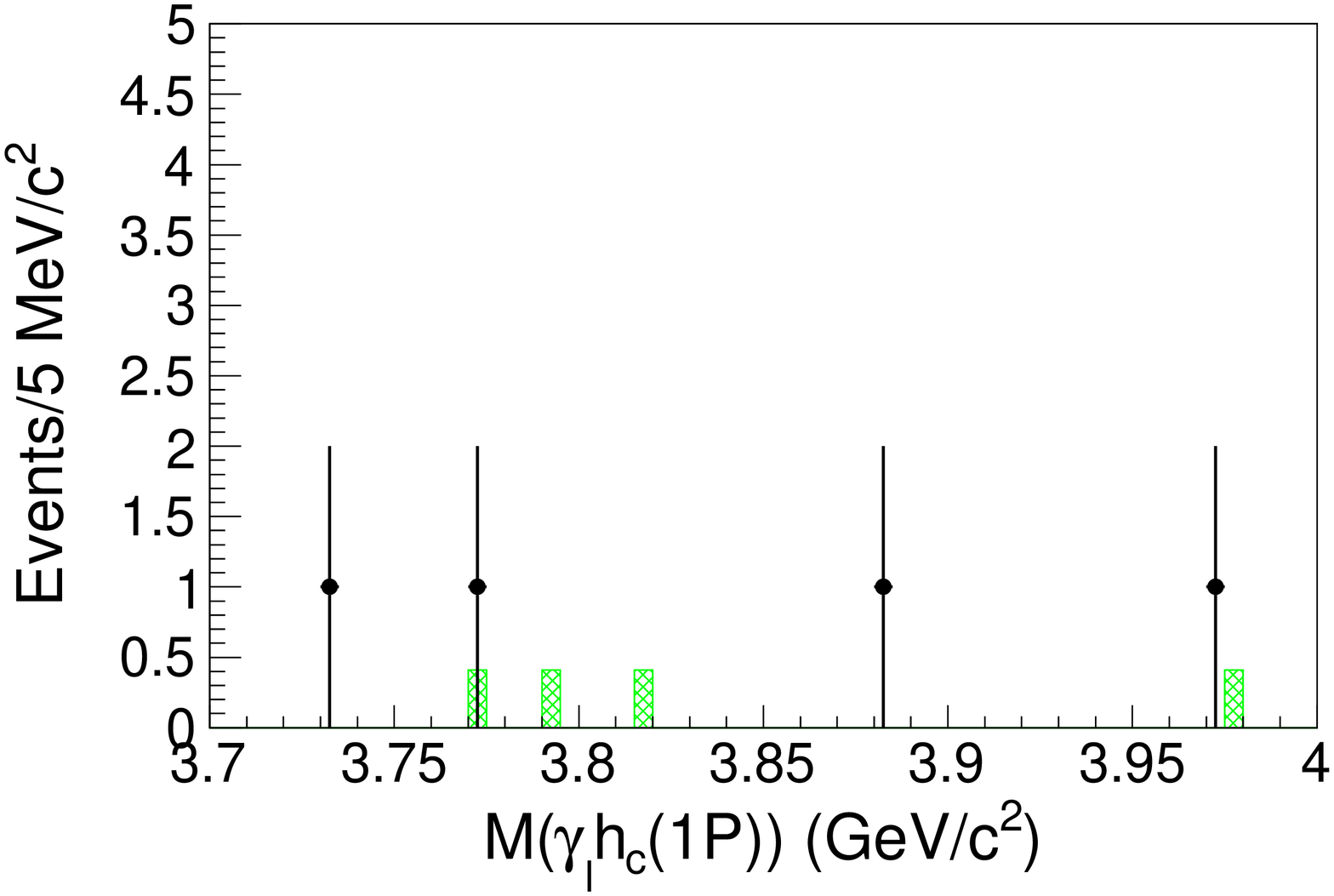}
\put(-25, -20){\large \bf (a)}
\hspace{0.1cm}
\includegraphics[width=3.7cm,angle=-90]{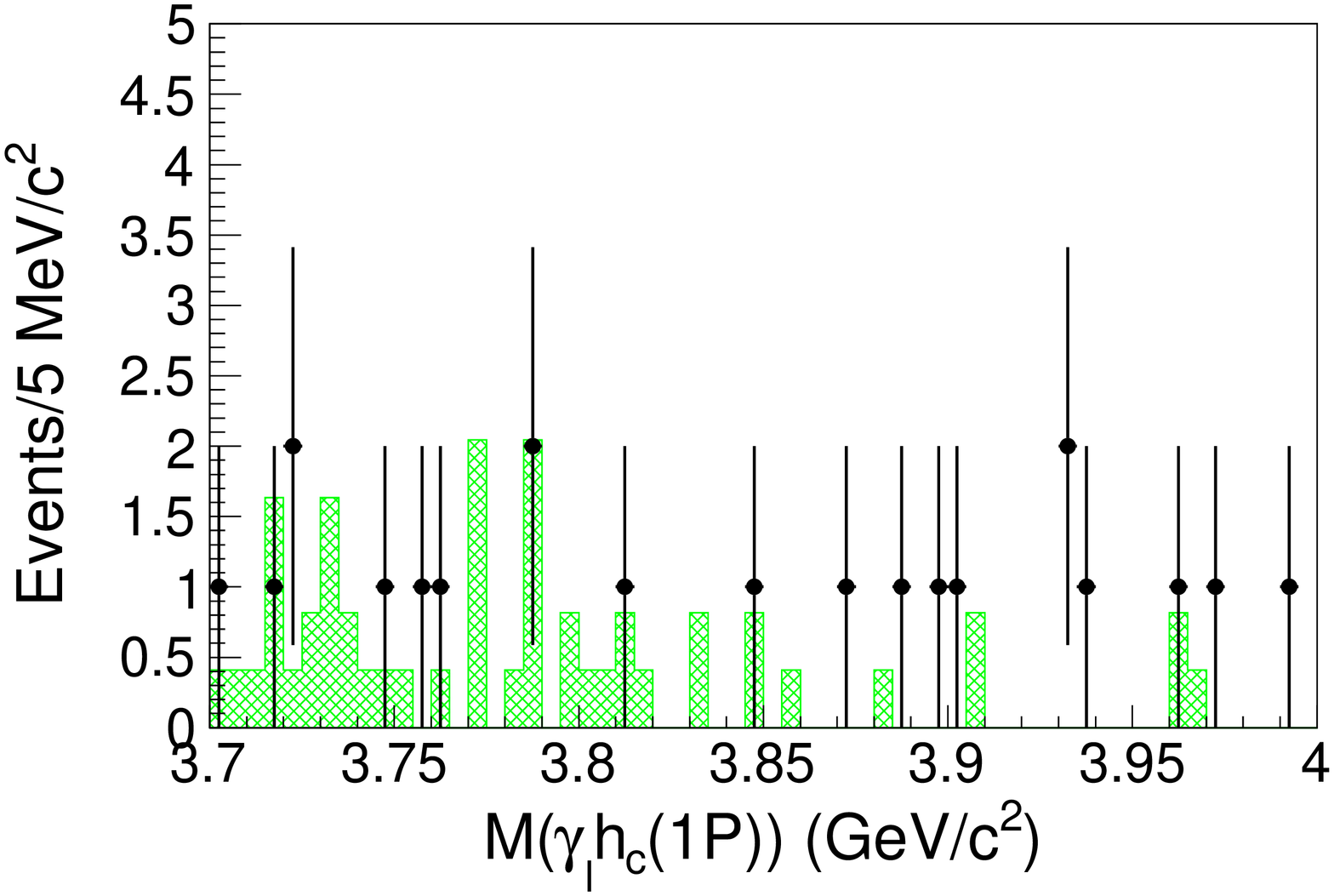}
\put(-25, -20){\large \bf (b)}
\hspace{0.10cm}
\includegraphics[width=3.7cm,angle=-90]{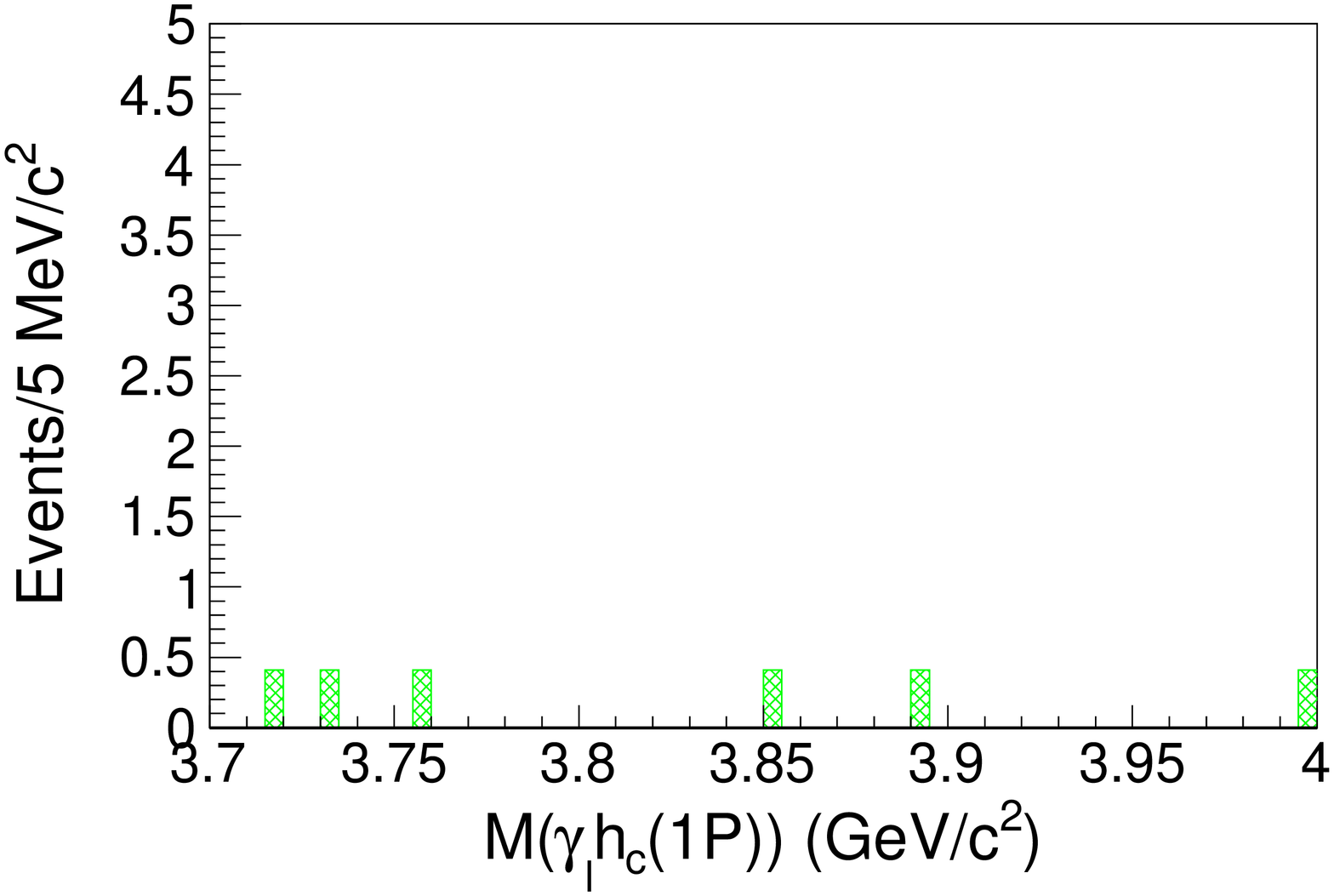}
\put(-25, -20){\large \bf (c)}
\caption{The invariant mass spectra for $\eta_{c2}(1D)$ candidates from $\sqrt{s}$ = (a) 10.52, (b) 10.58, and (c) 10.867 GeV data samples. The green cross-hatched histograms are from the normalized $h_c(1P)$ mass sidebands.}\label{etac2data}
\end{figure*}

The invariant mass distributions for $\eta_{c2}(1D)$ candidates from $\sqrt{s}$ = 10.52, 10.58, and 10.867 GeV data samples are shown in Fig.~\ref{etac2data}. No clear signals can be seen, and only a few candidates survived in the $\gamma_1 h_c(1P)$ mass spectra. The green cross-hatched histograms are from the normalized $h_c(1P)$ mass sidebands.

Since the number of selected signal candidate events is small, we obtain the 90\% C.L. upper limit of the signal yield ($N^{\rm UL}$) at each $\eta_{c2}(1D)$ mass point by using the frequentist approach~\cite{3873} implemented in the POLE (Poissonian limit estimator) program~\cite{012002}, where each mass region is selected to contain 95\% of the signal according to MC simulations, the number of signal candidate events is counted directly, and the number of expected background events is estimated from the normalized $h_c(1P)$ mass sidebands. The scan mass region is from 3.8 to 3.88 GeV/$c^2$ in steps of 4 MeV/$c^2$, which corresponds to the half mass resolution of $M(\gamma_1 h_c(1P))$. The systematic uncertainties discussed below have been taken into account in the POLE program~\cite{012002}.

\begin{figure*}[htbp]
\centering
\includegraphics[width=4cm,angle=-90]{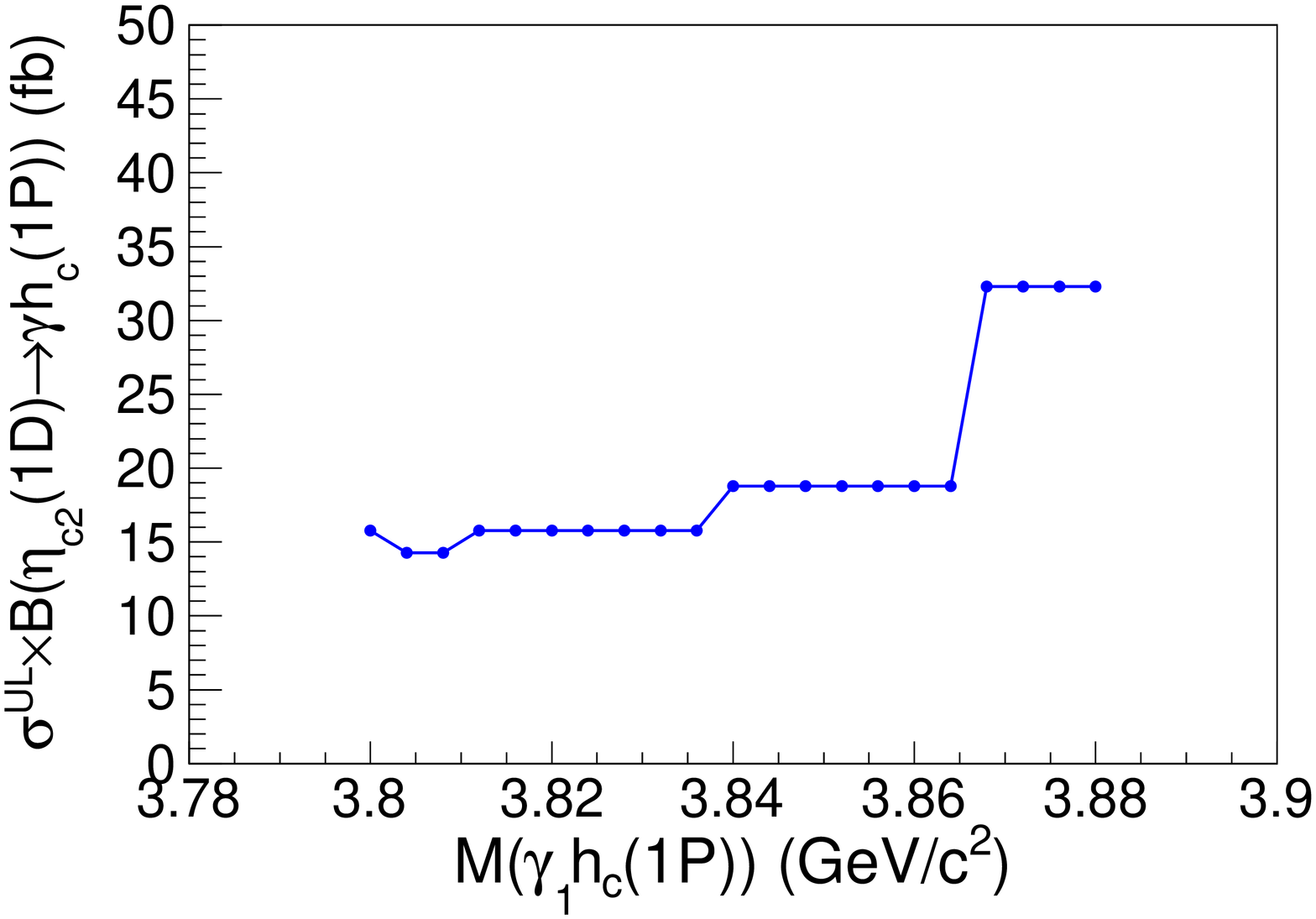}
\put(-25, -15){\large \bf (a)}
\hspace{0.2cm}
\includegraphics[width=4cm,angle=-90]{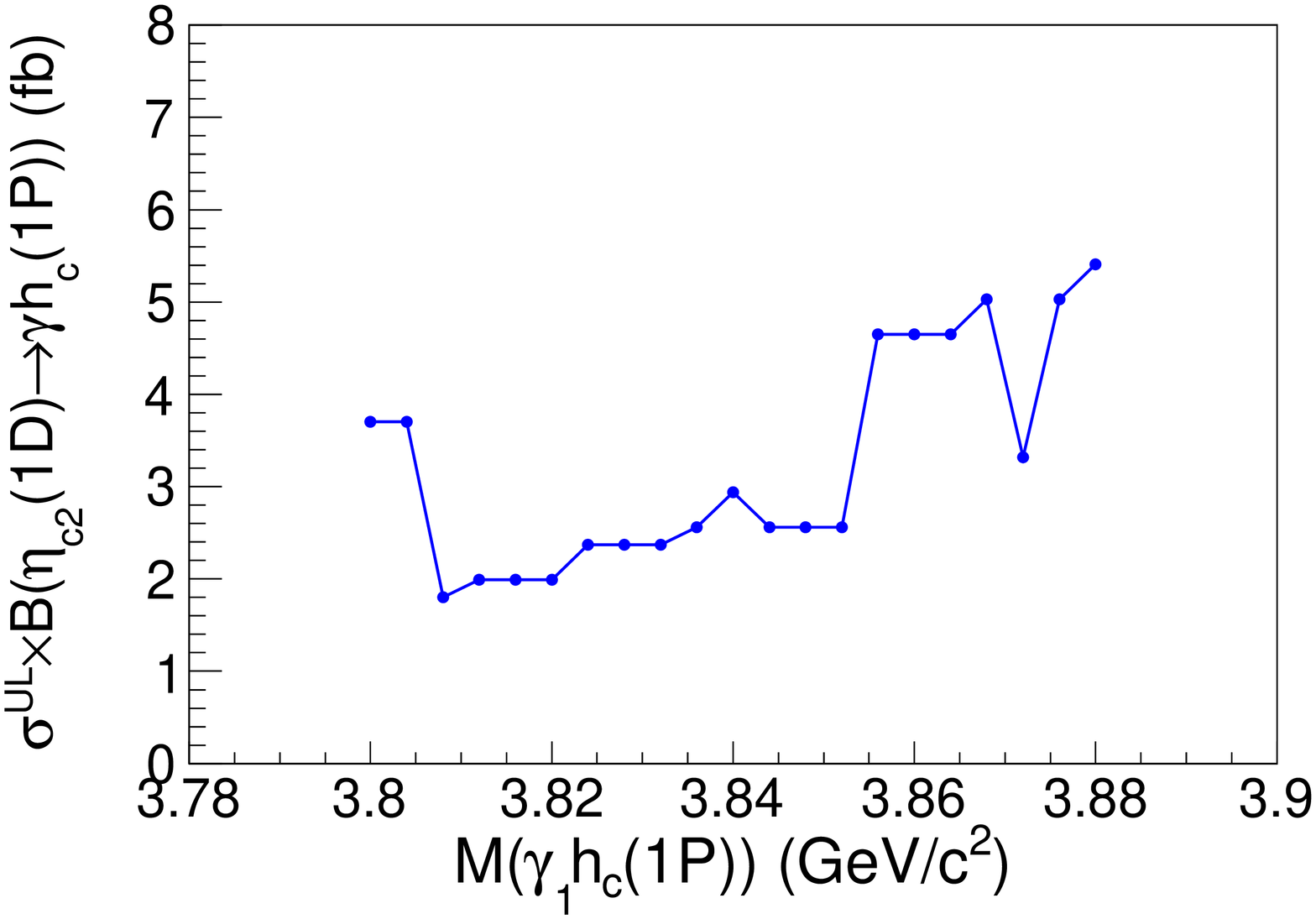}
\put(-25, -15){\large \bf (b)}
\vspace{0.2cm}

\includegraphics[width=4cm,angle=-90]{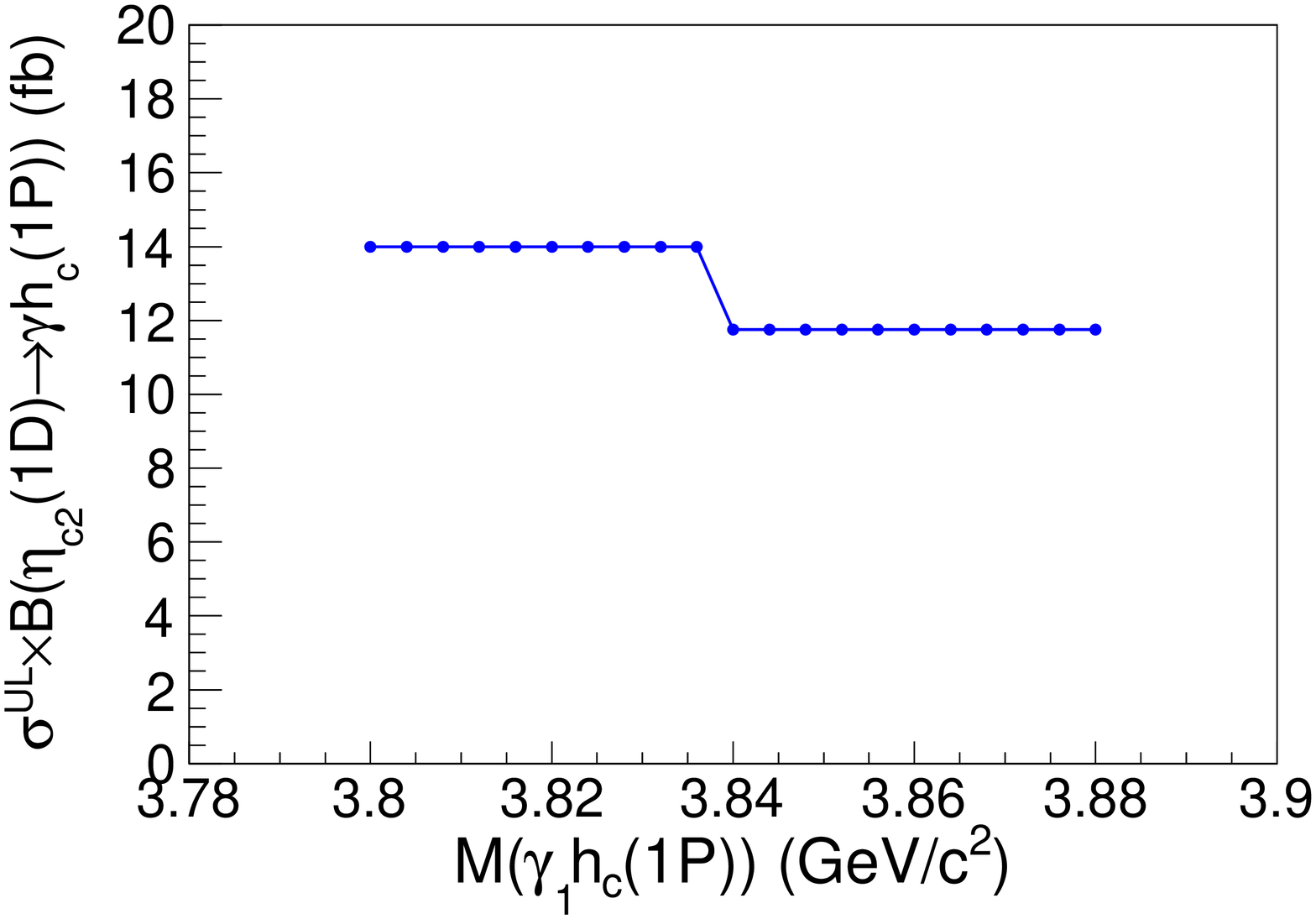}
\put(-25, -15){\large \bf (c)}
\hspace{0.2cm}
\includegraphics[width=4cm,angle=-90]{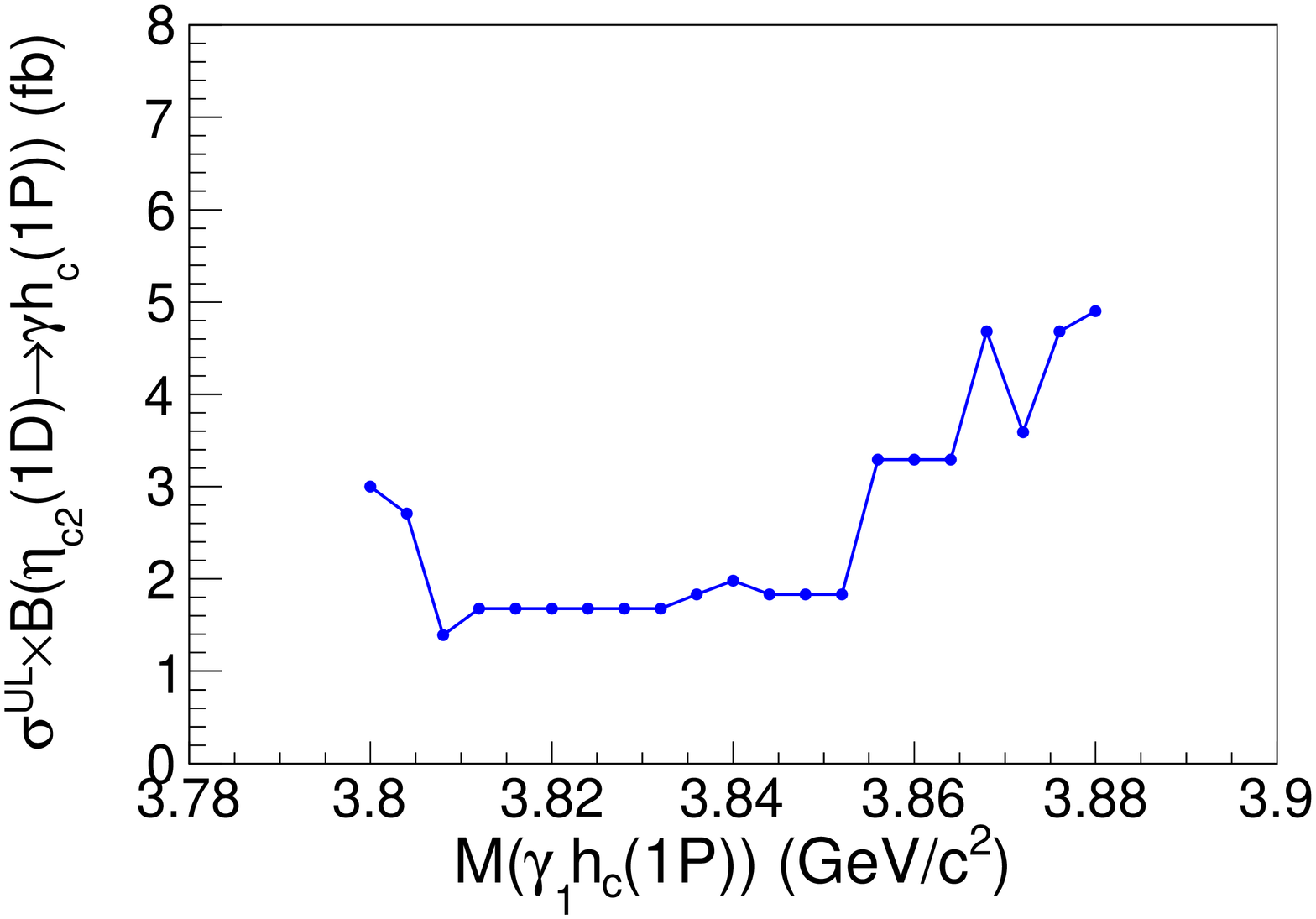}
\put(-25, -15){\large \bf (d)}
\caption{The upper limits on the product of the Born cross section for $e^+e^- \to \gamma\eta_{c2}(1D)$ and branching fraction for $\eta_{c2}(1D)\to\gamma h_c(1P)$ at $\sqrt{s}$ = (a) 10.52, (b) 10.58, (c) 10.867 GeV, and (d) from the combined data sample with systematic uncertainties included.}\label{ulsdata}
\end{figure*}

The upper limit on the product of the Born cross section for $e^+e^- \to \gamma\eta_{c2}(1D)$ and branching fraction for $\eta_{c2}(1D)\to\gamma h_c(1P)$ is calculated by
\begin{equation} \label{eq:1}
\begin{aligned}
\begin{split}
\sigma^{\rm UL}(e^+e^- \to \gamma\eta_{c2}(1D))\BR(\eta_{c2}(1D)\to\gamma h_c(1P)) = \\
\frac{N^{\rm UL}\times|1-\Pi|^2}{{\cal L}\times(1+\delta)_{\rm ISR}\times\Sigma_i\BR_i\varepsilon_i},~~~~~~~~~~~~~~
\end{split}
\end{aligned}
\end{equation}
where $N^{\rm UL}$ is the upper limit on the signal events in data, $|1-\Pi|^2$ is the vacuum polarization factor~\cite{585,CPC}, ${\cal L}$ is the integrated luminosity, $(1+\delta)_{\rm ISR}$ is the radiative correction factor~\cite{733}, and $\Sigma_i\BR_i\varepsilon_i$ is the sum over decay modes of the product of branching fractions and reconstruction efficiencies. The values of these variables (except $N^{\rm UL}$) are summarized in Table~\ref{t2}. The values of $\sigma^{\rm UL}(e^+e^- \to \gamma\eta_{c2}(1D))\BR(\eta_{c2}(1D)\to\gamma h_c(1P))$ at $\sqrt{s}$ = 10.52, 10.58, 10.867 GeV, and from the combined data sample are shown in Fig.~\ref{ulsdata}. The upper limit at 90\% C.L. on the product of the Born cross section for $e^+e^- \to \gamma\eta_{c2}(1D)$ and branching fraction for $\eta_{c2}(1D)\to\gamma h_c(1P)$ is 4.9 fb at $\sqrt{s}$ near 10.6 GeV.

\begin{table*}[htbp!]
\caption{The values used to determine the Born cross sections for $e^+e^- \to \gamma\eta_{c2}(1D)$ at $\sqrt{s}$ = 10.52, 10.58, 10.867 GeV, and from the  combined data sample, respectively.}\label{t2}
\vspace{0.2cm}
\centering
\begin{tabular}{c | c c c | c}
\hline\hline
$\sqrt{s}$ = & 10.52 GeV & 10.58 GeV & 10.867 GeV & Combined \\\hline
$|1-\Pi|^2$ & 0.931 & 0.930 & 0.929 & 0.930 \\
${\cal L}$ (fb) & 89.5 & 711 & 121.4 & 921.9\\
$(1+\delta)_{\rm ISR}$ & 0.679 & 0.679 & 0.677 & 0.679 \\
$\Sigma_i\BR_i\varepsilon_i$ & $(20.4\pm2.0)\times10^{-4}$ & $(20.3\pm2.0)\times10^{-4}$ &$(20.2\pm2.0)\times10^{-4}$ & $(20.3\pm2.0)\times10^{-4}$ \\
\hline\hline
\end{tabular}
\end{table*}

There are several sources of systematic uncertainty in the Born cross section measurements, including the detection efficiency, the uncertainty on the estimated signal efficiency due to limited MC statistics, the distribution of polar angle $\theta_\gamma$ for $e^+e^- \to \gamma\eta_{c2}(1D)$, intermediate state branching fractions, the energy dependence of the cross sections, and the integrated luminosity, which are listed in Table~\ref{t3}. The systematic uncertainty in detection efficiency is a final-state-dependent combined uncertainty for all the different types of particles detected, including tracking efficiency, PID, $K^0_S$ selection, and photon reconstruction.

\begin{table}[htbp]
\centering
\caption{Relative systematic uncertainties (\%) of the measurements of the Born cross sections for $e^+e^-\to \gamma \eta_{c2}(1D)$ at $\sqrt{s}$ = 10.52, 10.58, and 10.867$~\gev$, respectively.}\label{t3}
\begin{tabular}{c | c}
\hline\hline
Detection efficiency &7.0 \\
MC statistics &1.0 \\
~~~~~The distribution of $\theta_\gamma$~~~~~ &~~~~~13.7~~~~~ \\
Branching fractions &9.8 \\
Integrated luminosity &1.4 \\\hline
SUM & 18.3\\\hline\hline
\end{tabular}
\end{table}

Based on a study of $D^{*+} \to \pi^+ D^0, D^0 \to K_S^0 \pi^+ \pi^-$, the uncertainty in tracking efficiency is taken to be 0.35\% per track.~The uncertainties in PID for charged kaons and pions are determined to be 1.1\% and 1.0\%, respectively, based on a low-background sample of $D^{*+}$ decays.
The ratio of the efficiencies in $K_S^0$ selection between data and MC simulation is found to be $(97.9\pm 0.4\pm0.6)$\% in the decay chain of $D^{*+}\to \pi^+ D^0$, $D^0\to K^0_S\pi^+\pi^-$. We take 2.2\% as the systematic uncertainty due to $K_S^0$ selection. The uncertainty in the photon reconstruction is 2.0\% per photon, according to a study of radiative Bhabha events. The above individual uncertainties from different $\eta_c(1S)$ reconstructed modes are added linearly, weighted by the product of the detection efficiency and all secondary branching fractions ($\varepsilon_i\BR_i$). Assuming these
uncertainties are independent and adding them in quadrature, the final uncertainty related to the reconstruction efficiency is 7.0\%.

The statistical uncertainty in the determination of efficiency from signal MC samples is 1.0\%. For $\EE\to \gamma \eta_{c2}(1D)$, signal events are generated isotropically by default. Alternative angular distributions $(1 \pm \cos^2\theta_\gamma)$ are also generated, and the maximum difference in the detection efficiency between the alternatives and the default sample, 13.7\%, is taken as the systematic uncertainty.
Uncertainties from the $h_c(1P)$ and $\eta_c(1S)$ decay branching fractions are taken from the PDG~\cite{PDG}; the final uncertainties on intermediate state branching fractions are summed in quadrature over all the $\eta_c(1S)$ decay modes weighted by the product of the detection efficiency and secondary branching fractions ($\varepsilon_i\BR_i$).

Changing the $s$ dependence of the cross section from $n$ = 2 to $n$ = 1 or $n$ = 4, the differences of the radiative correction factor are very small ($<$ 1\%). Thus, this uncertainty is neglected. The total luminosity is determined to 1.4\% precision using wide-angle Bhabha scattering events. All the uncertainties are summarized in Table~\ref{t3} and, assuming all the sources are independent, summed in quadrature to give the total systematic uncertainty.

In summary, we search for $e^+e^-\to\gamma\eta_{c2}(1D)$ at $\sqrt{s}$ = 10.52, 10.58, and 10.867 GeV. We scan the $M(\gamma_1 h_c(1P))$ spectrum from 3.8 to 3.88 GeV/$c^2$ in steps of 4 MeV/$c^2$. No significant $\eta_{c2}(1D)$ signals are observed at any point.
The largest upper limit in the mass scans is regarded as the general upper limit. The upper limit at 90\% C.L. on the product of the Born cross section for $e^+e^- \to \gamma\eta_{c2}(1D)$ and branching fraction for $\eta_{c2}(1D)\to\gamma h_c(1P)$ is $\sigma(e^+e^- \to \gamma\eta_{c2}(1D))\BR(\eta_{c2}(1D)\to\gamma h_c(1P))$ $<$ 4.9 fb at $\sqrt{s}$ near 10.6 GeV. Taking $\BR(\eta_{c2}(1D)\to \gamma h_c(1P))$ $>$ 50\%~\cite{014027,054026,162002,014001}, the value of $\sigma(e^+e^- \to \gamma\eta_{c2}(1D))$ is smaller than 9.8 fb at 90\% C.L. at $\sqrt{s}$ near 10.6 GeV, which is consistent with the theoretical prediction of $\sigma(e^+e^- \to \gamma\eta_{c2}(1D))=1.5$ fb~\cite{14364}.

We thank the KEKB group for the excellent operation of the
accelerator; the KEK cryogenics group for the efficient
operation of the solenoid; and the KEK computer group, and the Pacific Northwest National
Laboratory (PNNL) Environmental Molecular Sciences Laboratory (EMSL)
computing group for strong computing support; and the National
Institute of Informatics, and Science Information NETwork 5 (SINET5) for
valuable network support.  We acknowledge support from
the Ministry of Education, Culture, Sports, Science, and
Technology (MEXT) of Japan, the Japan Society for the
Promotion of Science (JSPS), and the Tau-Lepton Physics
Research Center of Nagoya University;
the Australian Research Council including grants
DP180102629, 
DP170102389, 
DP170102204, 
DP150103061, 
FT130100303; 
Austrian Federal Ministry of Education, Science and Research (FWF) and
FWF Austrian Science Fund No.~P~31361-N36;
the National Natural Science Foundation of China under Contracts
No.~11435013,  
No.~11475187,  
No.~11521505,  
No.~11575017,  
No.~11675166,  
No.~11705209;  
No.~11761141009;
No.~11975076;
No.~12042509;
No.~12005040;
Key Research Program of Frontier Sciences, Chinese Academy of Sciences (CAS), Grant No.~QYZDJ-SSW-SLH011; 
the  CAS Center for Excellence in Particle Physics (CCEPP); 
the Shanghai Pujiang Program under Grant No.~18PJ1401000;  
the Shanghai Science and Technology Committee (STCSM) under Grant No.~19ZR1403000; 
the Ministry of Education, Youth and Sports of the Czech
Republic under Contract No.~LTT17020;
Horizon 2020 ERC Advanced Grant No.~884719 and ERC Starting Grant No.~947006 ``InterLeptons'' (European Union);
the Carl Zeiss Foundation, the Deutsche Forschungsgemeinschaft, the
Excellence Cluster Universe, and the VolkswagenStiftung;
the Department of Atomic Energy (Project Identification No. RTI 4002) and the Department of Science and Technology of India;
the Istituto Nazionale di Fisica Nucleare of Italy;
National Research Foundation (NRF) of Korea Grant
Nos.~2016R1\-D1A1B\-01010135, 2016R1\-D1A1B\-02012900, 2018R1\-A2B\-3003643,
2018R1\-A6A1A\-06024970, 2018R1\-D1A1B\-07047294, 2019K1\-A3A7A\-09033840,
2019R1\-I1A3A\-01058933;
Radiation Science Research Institute, Foreign Large-size Research Facility Application Supporting project, the Global Science Experimental Data Hub Center of the Korea Institute of Science and Technology Information and KREONET/GLORIAD;
the Polish Ministry of Science and Higher Education and
the National Science Center;
the Ministry of Science and Higher Education of the Russian Federation, Agreement 14.W03.31.0026, 
and the HSE University Basic Research Program, Moscow; 
University of Tabuk research grants
S-1440-0321, S-0256-1438, and S-0280-1439 (Saudi Arabia);
the Slovenian Research Agency Grant Nos. J1-9124 and P1-0135;
Ikerbasque, Basque Foundation for Science, Spain;
the Swiss National Science Foundation;
the Ministry of Education and the Ministry of Science and Technology of Taiwan;
and the United States Department of Energy and the National Science Foundation.

\renewcommand{\baselinestretch}{1.2}

\end{document}